\documentclass{article}
\usepackage{spconf,amsmath,graphicx}
\usepackage{amsmath,graphicx}
\usepackage{amsbsy}
\usepackage{epsfig}
\usepackage{float}
\usepackage{graphicx}
\usepackage{color}
\usepackage{url}
\usepackage{cite}
\usepackage{amssymb}
\usepackage{graphicx}
\usepackage{lipsum}
\usepackage{float}
\usepackage{cuted}
\usepackage{balance}
\usepackage{verbatim}
\usepackage{calc}  
\usepackage{enumitem}  

\usepackage{algorithmic}
\usepackage{algorithm2e}
\usepackage{subcaption}

\DeclareMathOperator*{\argmax}{arg\,max}
\DeclareMathOperator*{\argmin}{arg\,min}

\newcommand{\beq}{\begin{equation}}
\newcommand{\eeq}{\end{equation}}

\floatstyle{ruled}
\newfloat{algorithm}{tbp}{loa}
\providecommand{\algorithmname}{Algorithm}
\floatname{algorithm}{\protect\algorithmname}

\newcommand{\qedsymbol}{\hspace{\fill}\rule{1.5ex}{1.5ex}}

\makeatother

\ninept

\title{Topological Slepians: Maximally localized representations \\ of signals over simplicial complexes\vspace{-.1cm}}

\name{Claudio Battiloro, Paolo Di Lorenzo, and Sergio Barbarossa\vspace{-.2cm}}
\address{DIET Department, Sapienza University of Rome, Via Eudossiana 18, 00184, Rome, Italy\\ 
{E-mail: \{claudio.battiloro, paolo.dilorenzo, sergio.barbarossa\}@uniroma1.it}\vspace{-0.3cm}
}

\begin{document}

\maketitle

\begin{abstract}
This paper introduces topological Slepians, i.e., a novel class of signals defined over topological spaces (e.g., simplicial complexes) that are maximally concentrated on the topological domain (e.g., over a set of nodes, edges, triangles, etc.) and perfectly localized on the dual domain (e.g., a set of frequencies). These signals are obtained as the principal eigenvectors of a matrix built from proper localization operators acting over topology and frequency domains. Then, we suggest a principled procedure to build dictionaries of topological Slepians, which theoretically provide non-degenerate frames. Finally, we evaluate the effectiveness of the proposed topological Slepian dictionary in two applications, i.e., sparse signal representation and denoising of edge flows.
\end{abstract}


\begin{section}{Introduction}

In the last few years, there was a surge of interest in developing processing techniques for signals defined over irregular domains, which are not necessarily metric spaces. In particular, the field of graph signal processing (GSP) studies methodologies to analyze and process signals defined over the vertices of a graph \cite{shuman2013}, \cite{ortega2018graph}. To this aim, several graph operators (e.g. adjacency, Laplacian, etc.) have been introduced, thus leading to alternative designs of filters on graphs and graph Fourier transforms. The main feature of these processing tools is that they come to depend on the connectivity of the graph, which is encoded into the structure of the adopted graph operator. However, despite their overwhelming popularity, graph representations can only take into account {\it pairwise} relationships among data. 

In complex interconnected systems, the interactions often cannot be reduced to simple pairwise relationships, and graph representations might result incomplete and inefficient \cite{klamt2009hypergraphs,lambiotte2019networks, courtney2016generalized, giusti2016two, shen2018genome, benson2018simplicial, agarwal2006higher}. For instance, in biological networks, multi-way interactions among complex substances (such as genes, proteins, or metabolites) cannot be evoked using simply pairwise relationships \cite{lambiotte2019networks}; also, in the brain, groups of neurons typically activate at the same time \cite{giusti2016two}. These applications have sparked a renewed interest in extending GSP tools to incorporate multi-way relationships among data, thus leading to the emergent field of topological signal processing (TSP) \cite{barbarossa2020topological,schaub2021signal}. In this context, the seminal works \cite{barbarossa2020topological,schaub2021signal}
illustrated the benefits obtained by processing signals defined over simplicial complexes, which are specific examples of hyper-graphs with a rich algebraic description that can easily encode multi-way data. Then, several papers have given important contributions to TSP. For instance, the work in \cite{isufi2022scfilt} proposed FIR filters for signals defined over simplicial complexes, hinging on the Hodge decomposition, where the Fourier modes are eigenvectors of higher order combinatorial Laplacians \cite{schaub2020random}. The work in \cite{ji2022tsp} introduced generalized Laplacian for embedding simplicial complexes into traditional graphs. In \cite{coutino2020simplvolterra} the authors proposed self-driven graph Volterra models that can capture higher-order interactions among nodal observables available in networked data. Very recently, some works focused on processing signals defined over cell complexes, i.e., topological spaces not constrained to respect any inclusion property \cite{sardellitti2022cell, roddenberry2022cellsp}. Finally, deep neural architectures able to learn from data defined over topological spaces (e.g., simplicial or cell complexes) have been recently developed in \cite{bodnar2021weisfeiler,edgenetsIsufi,ebli2020simplicial,roddenberry2019hodgenet,giusti2021san,giusti2022cell}.

One of the fundamental problems in signal processing is to obtain a sparse representation of the signals of interest. Typically, the sparser is the representation, the better is the performance of several processing tasks such as, e.g., compression, denoising, sampling and recovery \cite{vetterlispbook}. In the context of TSP, a natural basis for signal representation is given by the topological Fourier modes \cite{schaub2020random}. However, often the signal of interest is highly localized over a portion of the topological domain, i.e., it is mostly concentrated over a set of nodes, edges, or triangles, while being at the same time highly (if not perfectly) localized over a specific set of frequencies. In such a case, Fourier modes might not lead to an efficient and sparse representation of localized topological signals, since they are usually non-sparse, i.e., not localized on sub-regions of the complex. To find localized spectral representations, the work in \cite{hodgelets2022roddenberry} proposed a family of wavelets for simplicial signals, respecting the Hodge decomposition and extending the graph wavelets from \cite{shuman2015graphwl}. 

In this work, moved by the necessity of finding efficient sparse representations of localized signals lying over simplicial complexes, we introduce topological Slepians, i.e., a class of signals that are maximally concentrated on the simplicial domain (e.g., over a set of nodes, edges, triangles, etc.) and perfectly localized on the dual domain (e.g., a set of frequencies). Due to their joint localization properties, they represent the simplicial counterpart of the prolate spheroidal wave functions introduced by Slepian and Pollack for continuous-time signals \cite{Slepian1961prolate}, and of the graph Slepians introduced in \cite{tsitsvero2016uncert,ville2017Slepiansbrain} for graph signals. These signals are obtained as the principal eigenvectors of a matrix built from localization operators defined over the topological space and its dual (frequency) domain. Then, we propose a principled way to build a dictionary having topological Slepians as atoms, identifying the theoretical conditions necessary to build a non-degenerate frame. Finally, we apply the proposed methodology to two cases of interest, namely, sparse signal representation and denoising of edge flows, illustrating how topological Slepians compare favourably with state-of-the-art techniques.
\end{section}

\vspace{-.2cm}
\begin{section}{Background}
\vspace{-.1cm}
In this section, we review basics of topological signal processing over simplical complexes that will be useful along the paper.\\
\noindent \textbf{Simplicial complex and signals.} Given a finite set of vertices $\mathcal{V}$, a $k$-simplex $\mathcal{H}_{k}$ is a subset of $\mathcal{V}$ with cardinality $k+1$, e.g., edges are $1$-simplices, and triangles are $2$-simplices. A face of $\mathcal{H}_{k}$ is a subset with cardinality $k$, and thus a $k$-simplex has $k+1$ faces. A coface of $\mathcal{H}_{k}$ is a $(k + 1)$-simplex
that includes $\mathcal{H}_{k}$ \cite{barbarossa2020topological, lim2020hodge}. If two simplices share a common face, then they are lower neighbours; if they share a common coface, they are upper neighbours \cite{yang2021finite}. A simplicial complex $\mathcal{X}_{k}$ of order $K$, is a collection of $k$-simplices $\mathcal{H}_{k}$, $k = 0, \ldots, K$ such that, for any $\mathcal{H}_{k} \in \mathcal{X}_{k}$, $\mathcal{H}_{k-1} \in \mathcal{X}_{k}$ if $\mathcal{H}_{k-1} \subset \mathcal{H}_{k}$ (inclusivity property). We denote the  set of $k$-simplices in $\mathcal{X}_{k}$ as  ${\cal D}_{k} := \{\mathcal{H}_{k}: \mathcal{H}_{k} \in \mathcal{X}_{k}\} $, with $|{\cal D}_{k}| = N_k$ and ${\cal D}_{k} \subseteq {\cal X}_{k}$. 

We are interested in processing signals defined over a simplicial complex. A $k$-simplicial signal is defined as a mapping from the set of all $k$-simplices contained in the complex to real numbers:
\begin{equation}
\mathbf{x}_{k}: {\cal D}_{k} \rightarrow \mathbb{R}, \,\,\quad k=0, 1, \ldots K.
\end{equation}
The order of the signal is one less the cardinality of the elements of ${\cal D}_{k}$. In most of the cases the focus is on complex $\mathcal{X}_{2}$ of order up to two, thus a set of vertices $\mathcal{V}$ with $|\mathcal{V}| = V$, a set of edges $\mathcal{E}$ with $|\mathcal{E}|=E$ and a set of triangles $\mathcal{T}$ with $|\mathcal{T}| = T$ are considered, resulting in ${\cal D}_{0}={\cal V}$ (simplices of order 0), ${\cal D}_{1}={\cal E}$ (simplices of order 1) and ${\cal D}_{2}={\cal T}$ (simplices of order 2).


\noindent\textbf{Algebraic representations.} The structure of a simplicial complex ${\cal X}_{k}$  is fully described by the set of its incidence matrices $\mathbf{B}_{k}$, $k=1, \ldots, K$, given a reference orientation. The entries of the incidence matrix $\mathbf{B}_{k}$ establish which $k$-simplices are incident to which $(k-1)$-simplices.  Denoting the fact that two simplex have the same orientation with ${H}_{k-1,i} \sim {H}_{k,j}$ and viceversa with ${H}_{k-1,i} \not\sim {H}_{k,j},$ the entries of $\mathbf{B}_{k}$ are defined as:
  \begin{equation} \label{inc_coeff}
  \big[\mathbf{B}_{k} \big]_{i,j}=\left\{\begin{array}{rll}
  \hspace{-.2cm}0, & \text{if} \; \mathcal{H}_{k-1,i} \not\subset \mathcal{H}_{k,j} \\
  \hspace{-.2cm}1,& \text{if} \; \mathcal{H}_{k-1,i} \subset \mathcal{H}_{k,j} \;  \text{and} \; \mathcal{H}_{k-1,i} \sim \mathcal{H}_{k,j}\\
  \hspace{-.2cm}-1,& \text{if} \; \mathcal{H}_{k-1,i} \subset \mathcal{H}_{k,j} \;  \text{and} \; \mathcal{H}_{k-1,i} \not\sim \mathcal{H}_{k,j}\\
  \end{array}\right. 
  \end{equation}
From the incidence information, we can build the high order combinatorial Laplacian matrices \cite{goldberg2002combinatorial} of order $k=0, \ldots, K$ as:
\begin{align}
&\mathbf{L}_{0}=\mathbf{B}_{1}\mathbf{B}_{1}^T,\label{Laplacian0}\\
&\mathbf{L}_{k}=\underbrace{\mathbf{B}_k^{T}\mathbf{B}_{k}}_{\mathbf{L}_k^{d}}+\underbrace{\mathbf{B}_{k+1}\mathbf{B}_{k+1}^T}_{\mathbf{L}_k^{u}}, \;\; k=1, \ldots, K-1, \label{Laplaciank}\\
&\mathbf{L}_{K}=\mathbf{B}_{K}^T\mathbf{B}_{K}.\label{LaplacianK}
\end{align}
All Laplacian matrices of intermediate order, i.e. $k=1, \ldots, K-1$, contain two terms: The first term $\mathbf{L}^{d}_k$, also known as  lower Laplacian, encodes the lower connectivity among $k$-order simplices; the second term $\mathbf{L}_k^{u}$, also known as upper Laplacian, encodes the upper connectivity among $k$-order simplices. For example, two edges are lower adjacent if they share a common vertex, whereas they are upper adjacent if they are faces of a common triangle.\\ 
\textbf{Hodge decomposition:} High order Laplacians admit a Hodge decomposition \cite{lim2020hodge}, such that the $k$-simplicial signal space can be decomposed as:
\begin{equation} \label{hodge_spaces}
\mathbb{R}^{N_{k}} = \textbf{im}\big(\mathbf{B}_{k}^T\big) \bigoplus \textbf{im}\big(\mathbf{B}_{k+1}\big) \bigoplus \textbf{ker}\big(\mathbf{L}_{k}\big),
\end{equation}
where $\bigoplus$ is the direct sum of vector spaces, and \textbf{ker}(·) and \textbf{im}(·) are the kernel and image spaces of a matrix, respectively. Thus, any signal $\mathbf{x}_{k}$ of order $k$ admits the following orthogonal decomposition:
\begin{equation}
\label{hodge_decomp}
    \mathbf{x}_{k}=\underbrace{\mathbf{B}_{k}^T\, \mathbf{x}_{k-1}}_{(a)} +\underbrace{\mathbf{B}_{k+1}\, \mathbf{x}_{k+1}}_{(b)} +\underbrace{\widetilde{\mathbf{x}}_{k}}_{(c)}  .
\end{equation}
To give an interpretation of (\ref{hodge_decomp}), let us consider the decomposition of edge flow signals $\mathbf{x}_{1}$, i.e., $k=1$. Then, the following holds \cite{barbarossa2020topological, yang2021simplicial}:
\begin{description}[leftmargin=!,labelwidth=\widthof{(a)}]
\item[(a)] Applying matrix $\mathbf{B}_{1}$ to an edge flow $\mathbf{x}_{1}$ computes its net flow (i.e. the difference between inflow and outflow) at each node, while
applying its adjoint $\mathbf{B}_{1}^T $ to a node signal $\mathbf{x}_{0}$ computes the gradients along the edges. We call $ \mathbf{B}_{1}^T\mathbf{x}_{0}$ the \textit{irrotational component} of $\mathbf{x}_{1}$ and $\textbf{im}(\mathbf{B}_{1}^T)$ the gradient space.
\item[(b)] The matrix $\mathbf{B}_{2}^T$ is a curl operator and applying it to an edge flow $\mathbf{x}_{1}$ computes the circulation over each triangle. Its adjoint $\mathbf{B}_{2}$ induces an edge flow $\mathbf{x}_{1}$ from a triangle signal $\mathbf{x}_{2}$. We call $\mathbf{B}_{2}\mathbf{x}_{2}$ the \textit{solenoidal component} of $\mathbf{x}_{1}$ and $\textbf{im}(\mathbf{B}_{2})$ the curl space.
\item[(c)] The remaining component $\widetilde{\mathbf{x}}_{1}$ is called the \textit{harmonic component} and  $\textbf{ker}(\mathbf{L}_{1})$ is called the harmonic space. Any edge flow $\widetilde{\mathbf{x}}_{1}$ has zero divergence and zero curl.
\end{description}
\noindent \textbf{Simplicial Fourier transform.} 
Simplicial signals of various order can be represented over the bases of the eigenvectors of the corresponding high order Laplacian matrices. Hence, using the eigendecomposition $\mathbf{L}_k=\mathbf{U}_k \mathbf{\Lambda}_k\mathbf{U}_k^T$, the Simplicial Fourier Transform (SFT) of order $k$ is defined as the projection of a $k$-order signal onto the eigenvectors of $\mathbf{L}_k$ \cite{barbarossa2020topological,yang2021finite}:
\begin{equation}
\label{k-GFT}
\widehat{\mathbf{x}}_k \triangleq \mathbf{U}_k^T\, \mathbf{x}_k.
\end{equation}
We refer to the eigenvalue domain of the SFT as the frequency domain. An interesting consequence of the Hodge decomposition in (\ref{hodge_decomp}) is that the eigenvectors belonging to $\textbf{im}(\mathbf{L}^{d}_k)$ are orthogonal to those belonging to  $\textbf{im}(\mathbf{L}^{u}_k)$, for all $k=1,\ldots,K-1$. Therefore, the eigenvectors of $\mathbf{L}_k$ are given by the union of the eigenvectors of $\mathbf{L}_k^u$ (spanning the curl sub-space), the eigenvectors of $\mathbf{L}_k^d$ (spanning the gradient sub-space), and the kernel of $\mathbf{L}_k$ (spanning the harmonic sub-space) \cite{yang2021finite}.
For the sake of simplicity, but without loss of generality, in the sequel we will focus on the processing of edge flow signals. Thus, we will denote $\mathbf{x}_{1}$ with $\mathbf{x}$, $\mathbf{U}_k$ with $\mathbf{U}$, $\mathbf{L}_{1}$ with $\mathbf{L}$, $\mathbf{L}_1^{d}$ with $\mathbf{L}^{d}$ and $\mathbf{L}_1^{u}$ with $\mathbf{L}^{u}$, such that $\mathbf{L} = \mathbf{L}^{d} + \mathbf{L}^{u}$. 
\end{section}

\vspace{-.2cm}
\begin{section}{Topological Slepians}
\vspace{-.1cm}

The eigenvectors of higher order Laplacians composing the simplicial Fourier basis are generally non-sparse, meaning that exploiting them for sparse representation of localized signals usually leads to poor performance \cite{hodgelets2022roddenberry}. Thus, in this section we introduce \textit{topological Slepians}, i.e., a novel class of simplicial signals that are maximally concentrated over a sub-set of edges, while being perfectly localized over the dual domain, i.e. a set of frequencies. Following the approach of \cite{tsitsvero2016uncert}, we introduce two localization operators acting onto an edge concentration set, say $\mathcal{S}$, and onto a frequency concentration set, say $\mathcal{F}$, respectively. The edge-limiting operator onto the edge set $\mathcal{S}$ is defined as the matrix $\mathbf{C}_{\mathcal{S}}\in\mathbb{R}^{E\times E}$ given by:
\begin{align}\label{edge_limiting_operator}
\mathbf{C}_{\mathcal{S}}={\rm diag}(\mathbf{1}_{\mathcal{S}}),
\end{align}
where $\mathbf{1}_{\mathcal{S}}\in\mathbb{R}^ {E}$ is a vector having ones in the index positions specified in $\mathcal{S}$, and zero otherwise; and ${\rm diag}(\mathbf{z})$ denotes a diagonal matrix having $\mathbf{z}$ on the diagonal. Clearly, from (\ref{edge_limiting_operator}), an edge signal $\mathbf{x}$ is perfectly localized onto the set $\mathcal{S}$ if $\mathbf{C}_{\mathcal{S}} \mathbf{x}=\mathbf{x}$. Similarly, the frequency limiting operator can be defined as:   
\begin{align}\label{frequency_limiting_operator}
\mathbf{B}_{\mathcal{F}}=\mathbf{U}\,{\rm diag}(\mathbf{1}_{\mathcal{F}})\,\mathbf{U}^T,
\end{align}
which represents an ideal band-pass filter over the frequency set $\mathcal{F}$. Clearly, an edge signal is perfectly localized over the bandwidth $\mathcal{F}$ if $\mathbf{B}_{\mathcal{F}} \mathbf{x}=\mathbf{x}$. The matrices in (\ref{edge_limiting_operator}) and (\ref{frequency_limiting_operator}) are projection operators, having maximum spectral radius equal to one. 

We define topological Slepians as the set of orthonormal vectors that are maximally concentrated over the edge set $\mathcal{S}$, and perfectly localized onto the bandwidth $\mathcal{F}$. Mathematically, topological Slepians are given by the solution of the following optimization problem:
\begin{align}
    \boldsymbol{\psi}_i = &\argmax_{\boldsymbol{\psi}_i}\; ||\mathbf{C}_\mathcal{S} \boldsymbol{\psi}_i||_2^2 \nonumber \\
    &\textrm{ subject to} \;\; \label{slep_prob}   ||\boldsymbol{\psi}_i|| = 1, \quad \mathbf{B}_\mathcal{F} \boldsymbol{\psi}_i = \boldsymbol{\psi}_i, \\
    & \; \qquad \qquad <\boldsymbol{\psi}_i,\boldsymbol{\psi}_j> = 0, \quad \hbox{$j = 1,\ldots,i-1$, if  $i > 1$,} \nonumber
\end{align}
for all $i=1,\ldots,E$. The solution of problem \eqref{slep_prob} is given by the following theorem, whose proof follows the same steps as in \cite{tsitsvero2016uncert} and is omitted due to lack of space.

\textit{Theorem 1.} The set of topological Slepians $\{\boldsymbol{\psi}_i\}_{i=1}^C$ with $C := \textrm{rank}\{\mathbf{B}_\mathcal{F}\mathbf{C}_\mathcal{S}\mathbf{B}_\mathcal{F}\}$ that solve problem (\ref{slep_prob}) is given by the eigenvectors of the matrix operator $\mathbf{B}_\mathcal{F}\mathbf{C}_\mathcal{F}\mathbf{B}_\mathcal{F}$, i.e.,
\begin{equation} \label{slep_sol}
    \mathbf{B}_\mathcal{F}\mathbf{C}_\mathcal{S}\mathbf{B}_\mathcal{F}\boldsymbol{\psi}_i = \lambda_i \boldsymbol{\psi}_i,
\end{equation}
where $\lambda_1 \geq  \lambda_2 \geq ... \geq \lambda_C > 0$. Furthermore, it holds
$<\boldsymbol{\psi}_i, \mathbf{C}_\mathcal{S}\boldsymbol{\psi}_j> = \lambda_j \delta_{i,j},$
where $\delta_{i,j}$ is the Kronecker delta. \qedsymbol

In Fig. \ref{fig::slep_ex}, we illustrate the first three topological Slepians maximally concentrated over the edge set in Fig. \ref{fig::slep_ex} (a), and bandlimited over the gradient sub-space (i.e., the set of frequency indexes associated with the eigenvectors of $\mathbf{L}^d$). Here, we consider topological Slepians for edge flows (i.e., $k=1$), but the approach can be similarly applied at any order $k=0,\ldots,K$, properly defining the operators (\ref{edge_limiting_operator}) and (\ref{frequency_limiting_operator}) over higher-order dual domains. In particular, the graph Slepians introduced in \cite{tsitsvero2016uncert} represent a very special case of our general formulation, obtained for $k=0$. Finally, the dual formulation of (\ref{slep_prob}), aimed at finding maximally localized signals over a set of frequencies, while being perfectly localized over an edge set, can be readily obtained exchanging the roles of $\mathbf{C}_\mathcal{S}$ and $\mathbf{B}_\mathcal{F}$ in (\ref{slep_prob}).
\end{section}

\vspace{-.1cm}
\begin{section}{Dictionary of Topological Slepians}
\vspace{-.1cm}

In this section we present a principled way to build overcomplete dictionaries of topological Slepians, relying on high order Laplacians in \eqref{Laplaciank} and the Hodge decomposition in \eqref{hodge_decomp}. Let $\boldsymbol{\Psi}_{\mathcal{S},\mathcal{F}}=\{\boldsymbol{\psi}_i\}_{i=1}^C$ be the set of topological Slepians obtained from (\ref{slep_sol}) considering the edge set $\mathcal{S}$ and the frequency set $\mathcal{F}$. In very general terms, a dictionary of topological Slepians has the following form:
\begin{equation}\label{slep_dic}
    \mathbf{D}_{\mathcal{C}} = \Big[\boldsymbol{\Psi}_{\mathcal{S}_1,\mathcal{F}_1},..., \boldsymbol{\Psi}_{\mathcal{S}_i,\mathcal{F}_i},...,\boldsymbol{\Psi}_{\mathcal{S}_M,\mathcal{F}_M} \Big],
\end{equation}
which collects $M$ sets of topological Slepians obtained from the pairs of concentration sets $\{\mathcal{S}_i,\mathcal{F}_i\}_{i=1}^M$. Therefore, building a dictionary of topological Slepians translates into properly choosing a sequence of concentration sets $\mathcal{C} = \{\mathcal{S}_i,\mathcal{F}_i\}_{i=1}^M$, i.e., a proper way of covering the edge set and the frequency domain. Let us initially assume that the kernel of the Laplacian $\mathbf{L}$ is empty, i.e., there are no harmonic components in (\ref{hodge_decomp}). Then, from (\ref{hodge_spaces}), the frequency indexes can be divided into two separate sets: (i) the \textit{irrotational band}, which corresponds to the set of eigenvalues of $\mathbf{L}^d$, say $\mathcal{F}^d$; and (ii) the \textit{solenoidal band}, which corresponds to the set of eigenvalues of $\mathbf{L}^u$, say $\mathcal{F}^u$. Thus, a very natural way to partition the frequency indexes is to choose $\mathcal{F}^d$ and $\mathcal{F}^u$ as frequency concentration sets, leading to irrotational and solenoidal Slepian components, respectively. Regarding the edge concentration sets, the Hodge Decomposition in \eqref{hodge_decomp} and the Laplacian structure in \eqref{Laplaciank} suggest us to define two distinct sequences of edge concentration sets: (i) $K^d$ sets $\{\mathcal{S}_i^d\}_{i=1}^{K^d}$ based on lower neighborhoods (encoded by $\mathbf{L}^d$); (ii) $K^u$ sets $\{\mathcal{S}_i^u\}_{i=1}^{K^u}$ based on upper neighborhoods (encoded by $\mathbf{L}^u$). Since the solenoidal and irrotational bandwidths are related to the eigenvectors of $\mathbf{L}^u$ and $\mathbf{L}^d$ [cf.  \eqref{hodge_decomp}], we propose to associate $\mathcal{F}^u$ to each upper concentration set $\mathcal{S}_i^u$, and $\mathcal{F}^d$ to each lower concentration set $\mathcal{S}_i^d$, such that the sequence $\mathcal{C}$ of concentration sets for the dictionary of topolgical Slepians in (\ref{slep_dic}) reads as (with a slight abuse of notation):
\begin{equation} \label{conc_sets}
\mathcal{C} = \{\mathcal{S}_i^u,\mathcal{F}^u\}_{i=1}^{K^u} \cup \{\mathcal{S}_i^d,\mathcal{F}^d\}_{i=1}^{K^d}.
\end{equation}
Finally, if the kernel of the Laplacian $\mathbf{L}$ is not empty, we propose to add to the dictionary $\mathbf{D}_{\mathcal{C}}$ also the eigenvectors $\overline{\mathbf{U}}= \{\mathbf{\overline{u}}_k\}_k$ of $\mathbf{L}$ associated with the zero eigenvalues. Therefore, given a sequence $\mathcal{C}$ of concentration sets built as in \eqref{conc_sets}, the corresponding dictionary of topological Slepians reads as:
\begin{equation} \label{slep_dic2}
    \mathbf{D}_{\mathcal{C}} = \Big[\boldsymbol{\Psi}_{\mathcal{S}_1^u,\mathcal{F}^u},...,\boldsymbol{\Psi}_{\mathcal{S}_{K^u}^u,\mathcal{F}^u} \boldsymbol{\Psi}_{\mathcal{S}_1^d,\mathcal{F}^d},...,\boldsymbol{\Psi}_{\mathcal{S}_{K^d}^d,\mathcal{F}^d}, \overline{\mathbf{U}}  \Big],
\end{equation}
where $M$ in (\ref{slep_dic}) is equal to $K^d+K^u+K^h$, $K^h = {\rm dim}(\textbf{ker}\big(\mathbf{L}\big))$. The size of the dictionary built in \eqref{slep_dic} can be controlled either tuning $K^d$ and $K^u$, or by taking just the top $\widetilde{K}$ Slepians per each pair of concentration sets. Clearly, the maximum possible number of Slepians depends on the chosen sets $\{\mathcal{S},\mathcal{F}\}$, because they affect the rank of the operator $\mathbf{B}_\mathcal{F}\mathbf{C}_\mathcal{S}\mathbf{B}_\mathcal{F}$ in \eqref{slep_sol}. Therefore, for a given $\widetilde{K}$ and pair of concentration sets $\{\mathcal{S},\mathcal{F}\}$, the number of topological Slepians will be $\min\{\widetilde{K},\textrm{rank}\{\mathbf{B}_\mathcal{F}\mathbf{C}_\mathcal{S}\mathbf{B}_\mathcal{F}\}\}$.

\begin{figure}[t]
\centering
\begin{subfigure}{0.49\columnwidth}
    \includegraphics[width=0.83\textwidth]{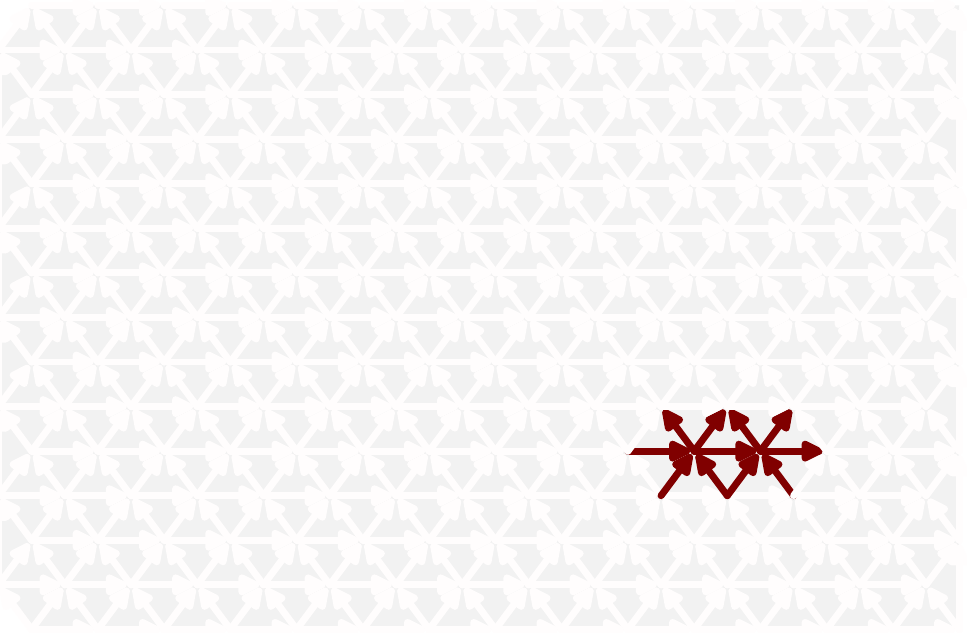}
    \caption{Example of edge concentration set}
\end{subfigure}
\hfill
\begin{subfigure}{0.49\columnwidth}
    \includegraphics[width=0.83\textwidth]{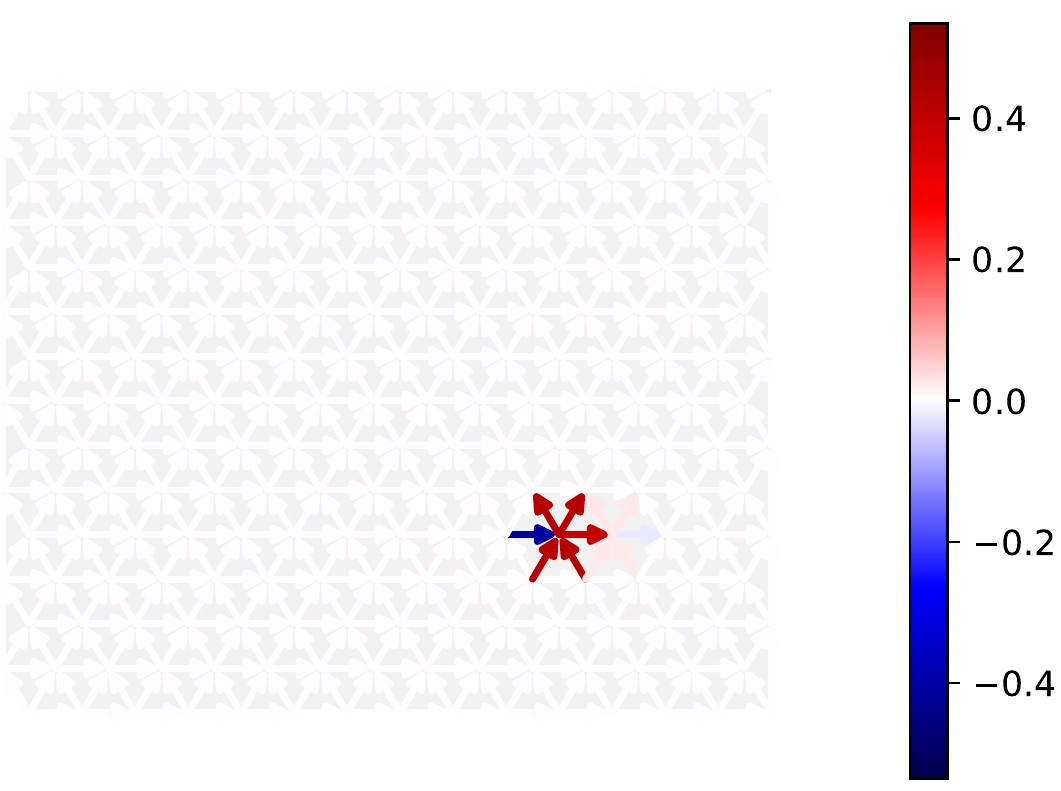}
    \caption{Topological Slepian nr. 1} 
\end{subfigure} 
\begin{subfigure}{0.49\columnwidth} 
    \includegraphics[width=0.83\textwidth]{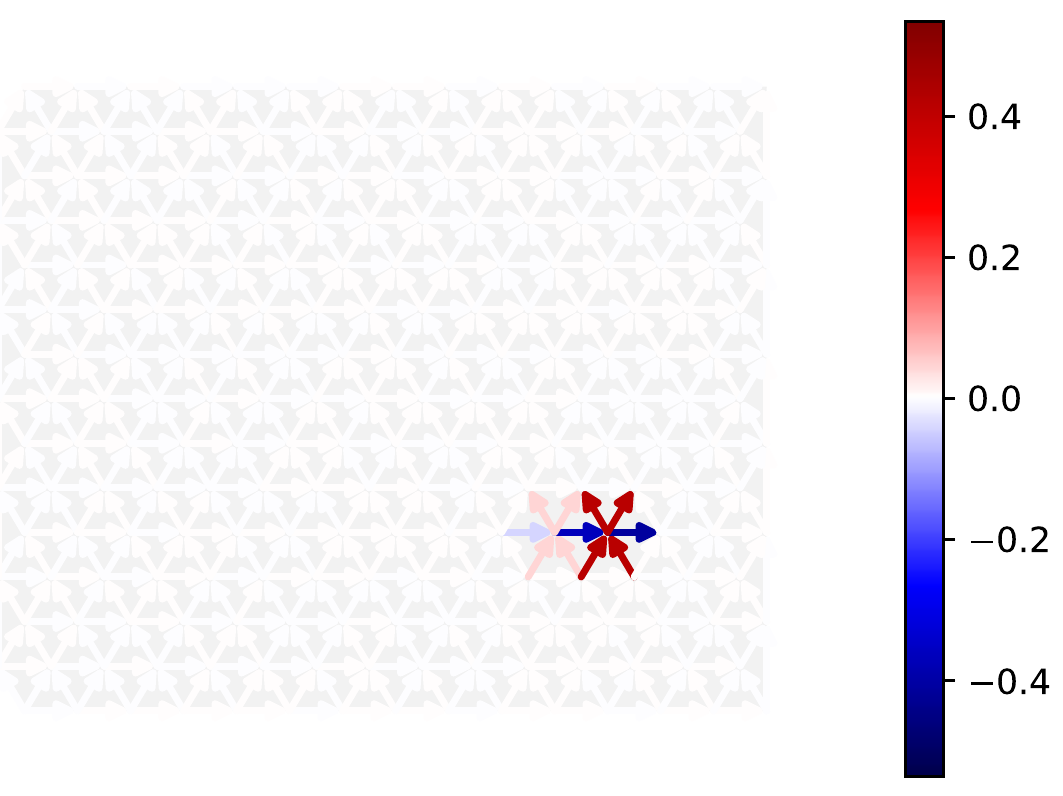} 
    \caption{Topological Slepian nr. 2} 
\end{subfigure}  
\hfill 
\begin{subfigure}{0.49\columnwidth} 
    \includegraphics[width=0.83\textwidth]{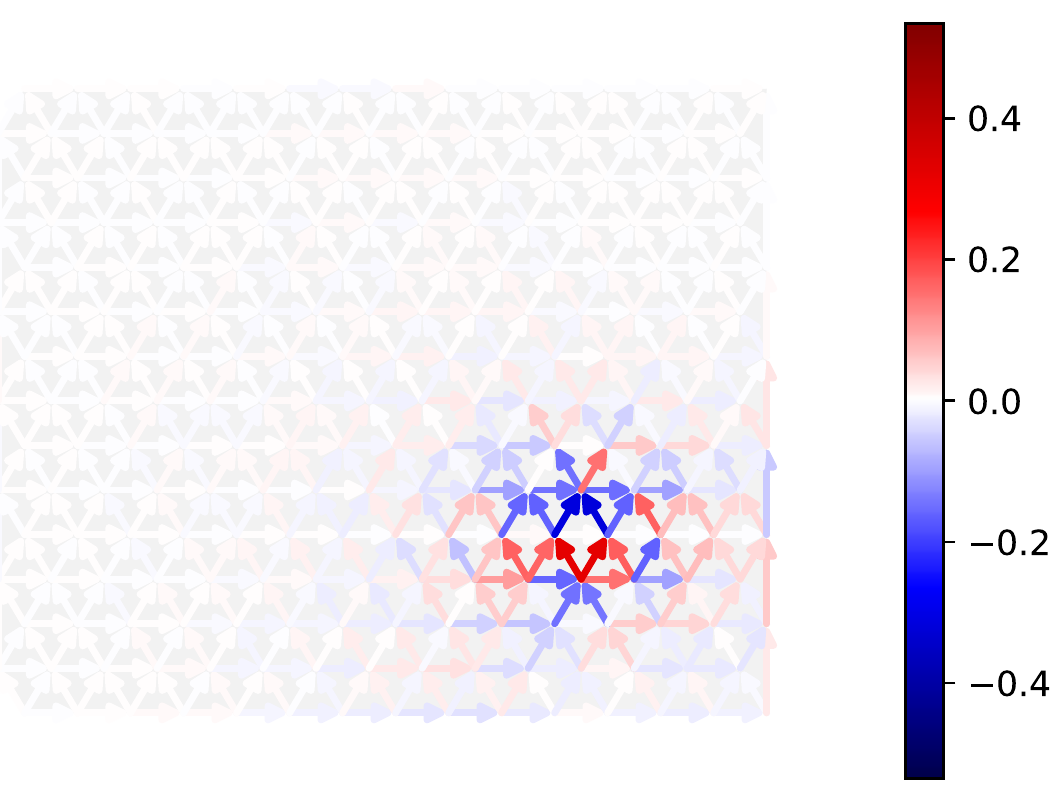} 
    \caption{Topological Slepian nr. 3} 
\end{subfigure} 
\caption{Example of topological Slepians on edge flows.}
\label{fig::slep_ex}
\end{figure}

\noindent \textbf{Frame bounds.} The dictionary $\mathbf{D}_{\mathcal{C}}\in\mathbb{R}^{E\times M}$ in \eqref{slep_dic2} is generally overcomplete, being composed by a set of $M>E$ non-orthogonal atoms. It is then of interest to investigate frame bounds on $\mathbf{D}_{\mathcal{C}}$. For a Hilbert space $\mathcal{H}$, a dictionary of vectors $\mathbf{D}_{\mathcal{C}}$ with at most countably
many elements is an $(A, B)$-frame if \cite{vetterlispbook}:
\begin{equation}\label{frame_def}
A||\mathbf{v}||_2^2 \leq \sum_{\boldsymbol{\psi} \in \mathbf{D}_{\mathcal{C}}} |<\boldsymbol{\psi},\mathbf{v}>|^2 \leq B||\mathbf{v}||_2^2, 
\end{equation}
for all $\mathbf{v} \in \mathcal{H}$, where $A$ and $B$ are called frame bounds, with $0\leq A \leq B < \infty$. If $A = B$, we say that $\mathbf{D}_{\mathcal{C}}$ is a tight frame; if $A=0$, $\mathbf{D}_{\mathcal{C}}$ is said to be a degenerate frame. Let $\boldsymbol{\Psi}^u=\{\boldsymbol{\Psi}_{\mathcal{S}_i^u}\}_{i=1}^{K^u}$ and $\boldsymbol{\Psi}^d=\{\boldsymbol{\Psi}_{\mathcal{S}_i^d}\}_{i=1}^{K^d}$ be the matrices collecting all the upper and lower Slepians, respectively. The following Theorem shows that, under mild conditions on the choice of the edge concentration sets, a Slepian dictionary as in \eqref{slep_dic} forms an non-degenarate $(A, B)$-frame.

\textit{Theorem 2.} Let $\mathbf{D}_{\mathcal{C}}$ be a dictionary of topological Slepians as in \eqref{slep_dic}. Let the following properties hold:\vspace{-.15cm}
\begin{description}
\item[\quad A1 (Lower completeness):]
The lower localization sets $\{\mathcal{S}_i^d\}_{i=1}^{K^d}$ are chosen such that    ${\rm colspan}\{\boldsymbol{\Psi}^d\}= \textbf{im}(\mathbf{L}^d)$. \vspace{-.15cm}
\item[\quad A2 (Upper completeness):] The upper localization sets $\{\mathcal{S}_i^u\}_{i=1}^{K^d}$ are chosen such that
    ${\rm colspan}\{\boldsymbol{\Psi}^u\}= \textbf{im}(\mathbf{L}^u).$ \vspace{-.1cm}
\end{description}
Then, $\mathbf{D}_{\mathcal{C}}$ is a non-degenerate $(A,B)$-frame for $\mathbb{R}^E$ with $A = \lambda_{min}(\sum_{\boldsymbol{\psi} \in \mathbf{D_\mathcal{C}}}\boldsymbol{\psi}\boldsymbol{\psi}^T)>0$ and $B=K^d+K^u+K^h$. \smallskip 

\textit{Proof.} See Appendix A.   \qedsymbol

\vspace{.1cm}
\noindent In Theorem 2, A1 requires that the set of lower Slepians spans the gradient space, i.e., $\textbf{im}(\mathbf{L}^d)$. Similarly, A2 assumes that the set of upper Slepians spans the curl space, i.e., $\textbf{im}(\mathbf{L}^u)$. These conditions are sufficient for the claim of the theorem, and subsume some necessary conditions for the choice of the localization sets. The first one is related to coverage, i.e., the union of the localization sets defining the lower and upper Slepians must cover the whole edge set. Mathematically, we need $\bigcup_{i=1}^{K^d} \mathcal{S}_i^d=\bigcup_{i=1}^{K^u} \mathcal{S}_i^u=\mathcal{E}$. The second condition is related to the number of elements in the dictionary that must satisy $K^d \geq |\mathcal{F}^d|$ and $K^u \geq |\mathcal{F}^u|$, which is necessary to find a set of lower and upper Slepians that span  $|\mathcal{F}^d|$ and $|\mathcal{F}^u|$ dimensional subspaces, i.e., the gradient and the curl spaces, respectively. 
Interestingly, the conditions of Theorem 2 can be easily satisfied using very simple heuristic procedures for the selection of the localization sets $\{\mathcal{S}_i^d\}_{i=1}^{K^d}$ and  $\{\mathcal{S}_j^u\}_{j=1}^{K^u}$. For instance, an easy and low-complexity example is building a dictionary by choosing the upper and lower edge concentration sets as the $1$-hop upper and lower neighborhoods of each edge including the edge itself, respectively. In the following numerical results, we will exploit this simple procedure, but the choice of the localization sets is an interesting open problem that could be addressed resorting to hierarchical higher order spectral clustering techniques, with the objective of finding tighter frames.

\end{section}

\begin{figure}[t]
    \centering
    \includegraphics[scale = 0.21]{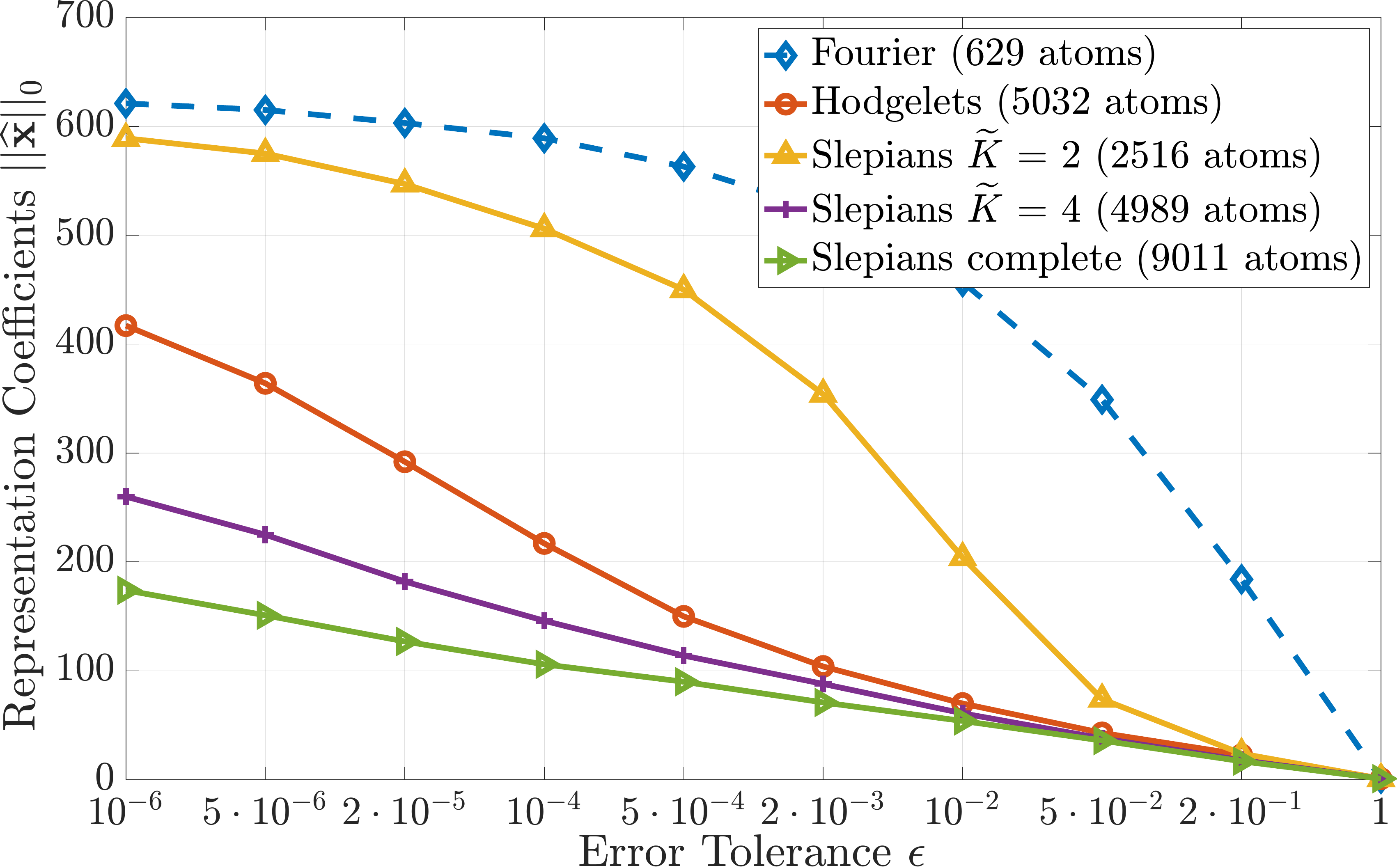}
    \caption{Sparsity vs error tolerance.}
    \label{fig::spars_eps}
\end{figure}

\vspace{-.15cm}
\begin{section}{Numerical Results}
\vspace{-.15cm}
We assess the performance of topological Slepians dictionaries on two tasks, i.e., sparse signal representation, and edge flow denoising (code at https://github.com/clabat9/Topological-Slepians). We rely on the experimental setting of \cite{hodgelets2022roddenberry}, considering the localized vector field on $[-2,2]^2$ given by 
 $   F(x,y) = [\cos(x+y),\sin(x-y)] \mathbb{I}\Big((x,y) \in B_1 \cup B_2 \Big),$
where $\mathbb{I}$ is the indicator function, $B_1$ and $B_2$ are closed balls of radius 0.7 centered
at $(\pm \pi/4,\pm \pi/4)$, respectively. The vector field is then discretized with a hexagonal grid, and a simplicial complex is built ($V$ = 225, $E$ = 629, $T$ = 405) by treating each hexagon as a node, with edges for each pair of hexagons that share a side, and triangles for each set of three hexagons that share a corner. Finally, the vector field is converted to an edge signal $\mathbf{x} \in \mathbb{R}^E$ by taking the flow on each edge to be the total flow perpendicular to the corresponding side between hexagons; $\mathbf{x}$ is normalized to have unit norm. 
The dictionary of topological Slepians is built as in \eqref{slep_dic2}, considering $2E$ edge concentration sets, defined as the upper and the lower $1$-hop neighborhood of each edge, in order to satisfy the conditions of Theorem 2. No harmonic component is present.


\noindent\textbf{Sparse signal representation -} We first evaluate how sparse is the representation of $\mathbf{x}$ on the proposed dictionary. We employ orthogonal matching pursuit (OMP), which seeks the sparsest linear combination of the Slepians in the dictionary by greedily solving \cite{omp2007}: 
\begin{align}\label{OMP_problem}
    \widehat{\mathbf{x}} = \argmin ||\widehat{\mathbf{x}}||_0 \quad \hbox{subject to} \quad \| \mathbf{x} - \mathbf{D}_\mathcal{C}\widehat{\mathbf{x}}\|^2 \leq \epsilon.
\end{align}
In Fig. \ref{fig::spars_eps}, we illustrate the $\ell_0$ norm of the obtained $\widehat{\mathbf{x}}$ versus the error tolerance $\epsilon$, comparing the Fourier basis, the Hodgelets \cite{hodgelets2022roddenberry}, and the topological Slepians dictionaries obtained for some values of $\widetilde{K}$. As we can see from Fig. \ref{fig::spars_eps}, the proposed Slepians dictionary shows superior performance with respect to the Hodgelets \cite{hodgelets2022roddenberry}, being able to find sparser representations even for smaller dictionaries, i.e., for $\widetilde{K}=4$. The complete Slepians dictionary leads to the best performance, at the cost of a larger dictionary size and complexity. 

\begin{figure}[t]
    \centering
    \includegraphics[scale = 0.21]{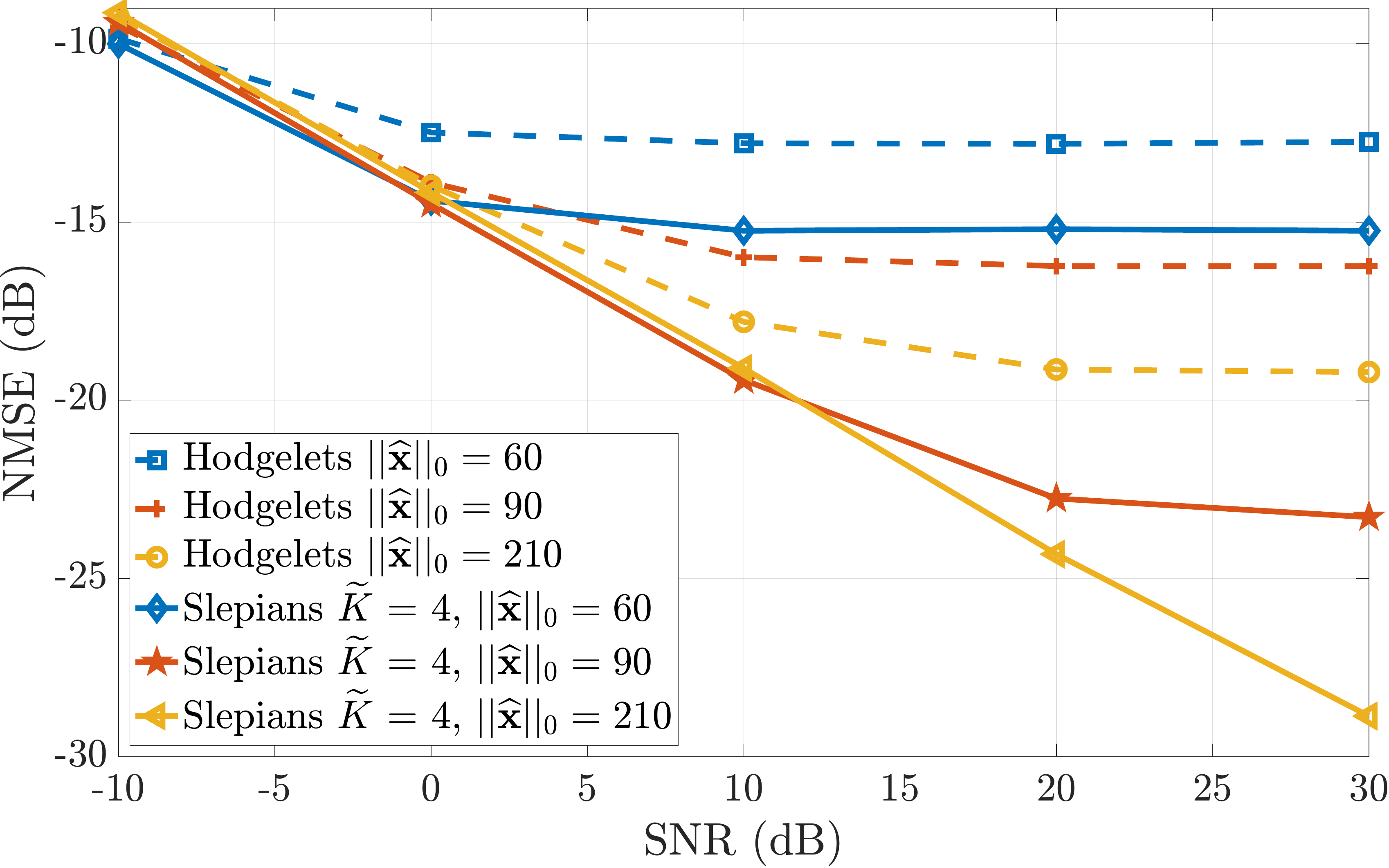}
    \caption{NMSE vs SNR, for different denoising strategies.}
    \label{fig::nmse_snr}
\end{figure}
\noindent\textbf{Edge flow denoising -} As a second experiment, we evaluate the effectiveness of topological Slepians diactionaries on a task of edge flow denoising. In particular, we take the same edge signal $\mathbf{x}$ of the previous experiment and we add independent Gaussian noise to it, obtaining 
$\widetilde{\mathbf{x}} =\mathbf{x} + \mathbf{n}$, where $\mathbf{n}  \overset{i.i.d.}{\sim} \mathcal{N}(0,\sigma^2)$. 
Now, we exploit again OMP, but switching the objective function and the constraint of \eqref{OMP_problem}, so as to minimizee the mean squared error, under a sparsity constraint on the $\ell_0$-norm of $\widehat{\mathbf{x}}$, with the goal of  evaluating how good is the obtained representation $\mathbf{D}_\mathcal{C}\widehat{\mathbf{x}}$ in estimating the clean signal  $\mathbf{x}$. In Fig. \ref{fig::nmse_snr}, we show the Normalized Mean Square error $||\mathbf{x}-\mathbf{D}_\mathcal{C}\widehat{\mathbf{x}}||^2$ versus the Signal to Noise Ratio (SNR), averaged over 100 independent simulations, comparing  Hodgelets and topological Slepians dictionary (with $\widetilde{K}=4$) obtained for three values of $||\widehat{\mathbf{x}}||_0$; the SNR is defined as $\textrm{SNR} = ||\mathbf{x}||^2/(\sigma^2 \cdot E) = 1/(\sigma^2 \cdot E)$. From Fig. \ref{fig::nmse_snr} it is evident that topological Slepians leads to better estimation performance, for any value of SNR and $||\widehat{\mathbf{x}}||_0$.


\end{section}

\vspace{-.15cm}
\begin{section}{Conclusions}
\vspace{-.15cm}

In this paper we introduced topological Slepians, a novel class of signals that are maximally localized over a simplicial complex, while being perfectly localized onto a dual (frequency) domain. Hinging on Hodge decomposition and exploiting the structure of high-order Laplacians, we have proposed a principled method to build dictionaries of topological Slepians and we derived their frame bounds. Finally, we assessed their effectiveness in sparse signal representation and edge flow denoising when compared to state of the art approaches available in the literature.
\end{section}

\balance
\bibliographystyle{IEEEbib}

\bibliography{biblio.bib}

\appendix
\section{Proof of Theorem 2}
 Let's start showing that $B = K^d+K^u+K^h$. We first notice that:
 \begin{align}\label{slep_proof_sum}
     \sum_{\boldsymbol{\psi} \in \mathbf{D_\mathcal{C}}}|<\boldsymbol{\psi},\mathbf{v}>|^2 &=  \sum_{k=1}^{K^d}\sum_{c=1}^{\widetilde{K}}|<\boldsymbol{\psi}^d_{k,c},\mathbf{v}>|^2 +  \nonumber \\
     &\hspace{-2cm}+ \sum_{k=1}^{K^u}\sum_{c=1}^{\widetilde{K}}|<\boldsymbol{\psi}^u_{k,c},\mathbf{v}>|^2+ \sum_{k=1}^{K^h}|<\mathbf{\widetilde{u}}_k,\mathbf{v}>|^2
\end{align}
 From Bessel's inequality, we have
 $\sum_{c=1}^{\widetilde{K}}|<\boldsymbol{\psi}^d_{k,c},\mathbf{v}>|^2 \leq \|\mathbf{v}\|^2$, and $\sum_{c=1}^{\widetilde{K}}|<\boldsymbol{\psi}^d_{k,c},\mathbf{v}>|^2 \leq \|\mathbf{v}\|^2$. Plugging such inequalities in (\ref{slep_proof_sum}), and exploiting $\|\mathbf{\widetilde{u}}_k\|=1$ for $k=1,\ldots,K^h$,  we obtain that each term of the form $\sum_{c=1}^{\widetilde{K}}|<\boldsymbol{\psi}^u_{k,c},\mathbf{v}>|^2$ is smaller than $||\mathbf{v}||_2^2$ and so is $\sum_{k=1}^{K^h}|<\mathbf{\widetilde{u}},\mathbf{v}>|^2$. Thus, from (\ref{slep_proof_sum}), we have:
 \begin{align}
    \sum_{\boldsymbol{\psi} \in \mathbf{D_\mathcal{C}}}|\leq \boldsymbol{\psi},\mathbf{v}>|^2 < (K^d+K^u+K^h)||\mathbf{v}||_2^2.
\end{align}
Therefore $B = K^d+K^u+K^h$. 

Now we show that $A = \lambda_{\textrm{min}}(\sum_{\boldsymbol{\psi} \in \mathbf{D_\mathcal{C}}}\boldsymbol{\psi}\boldsymbol{\psi}^T)$ and that $A>0$. 
 The first claim comes directly from the Rayleigh-Ritz inequality:
 \begin{align}\label{RR_inequality}
    \sum_{\boldsymbol{\psi} \in \mathbf{D_\mathcal{C}}}|<\boldsymbol{\psi},\mathbf{v}>|^2 & \,= \,\mathbf{v}^T \left(\textstyle\sum_{\boldsymbol{\psi} \in \mathbf{D_\mathcal{C}}}\boldsymbol{\psi}\boldsymbol{\psi}^T \right)\mathbf{v}\nonumber\\
     &\,\geq\, \lambda_{\textrm{min}}\left(\textstyle\sum_{\boldsymbol{\psi} \in \mathbf{D_\mathcal{C}}}\boldsymbol{\psi}\boldsymbol{\psi}^T\right) \|\mathbf{v}\|^2.
\end{align}
 Then, $A>0$ holds true if $\sum_{\boldsymbol{\psi} \in \mathbf{D_\mathcal{C}}}\boldsymbol{\psi}\boldsymbol{\psi}^T$ is full rank. From (\ref{RR_inequality}) and (\ref{slep_proof_sum}), we have
 \begin{align}
     \sum_{\boldsymbol{\psi} \in \mathbf{D_\mathcal{C}}}\boldsymbol{\psi}\boldsymbol{\psi}^T&= \underbrace{\sum_{k=1}^{K^u}\sum_{c=1}^{\widetilde{K}}\boldsymbol{\psi}^d_{k,c}\boldsymbol{\psi}^{d\,T}_{k,c}}_{\mathbf{G}_c^d} + \underbrace{\sum_{k=1}^{K^u}\sum_{c=1}^{\widetilde{K}}\boldsymbol{\psi}^u_{k,c}\boldsymbol{\psi}^{u\,T}_{k,c}}_{\mathbf{G}_c^u} + \underbrace{\sum_{k=1}^{K^h}\widetilde{\mathbf{u}}_k\widetilde{\mathbf{u}}_k^T}_{\mathbf{G}_c^h}.\nonumber
 \end{align}
 From assumptions A1 and A2, $\mathbf{G}_c^d$ and $\mathbf{G}_c^u$ have rank equal to $|\mathcal{F}^d|$ and $|\mathcal{F}^u|$, respectively, whereas $\textrm{rank}(\mathbf{G}_c^h) = K^h$. Then, exploiting the orthogonality between upper, lower, and harmonic Slepians, we obtain $\textrm{rank}(\sum_{\boldsymbol{\psi} \in \mathbf{D_\mathcal{C}}}\boldsymbol{\psi}\boldsymbol{\psi}^T) = |\mathcal{F}^d|+|\mathcal{F}^u|+K^h= E$, and thus $A=\lambda_{\textrm{min}}(\sum_{\boldsymbol{\psi} \in \mathbf{D_\mathcal{C}}}\boldsymbol{\psi}\boldsymbol{\psi}^T)>0$.

\end{document}